\documentstyle{ws-p8-50x6-00}

\newcommand{\drawsquare}[2]{\hbox{%
\rule{#2pt}{#1pt}\hskip-#2pt
\rule{#1pt}{#2pt}\hskip-#1pt
\rule[#1pt]{#1pt}{#2pt}}\rule[#1pt]{#2pt}{#2pt}\hskip-#2pt
\rule{#2pt}{#1pt}}
\newcommand{\Yfund}{\raisebox{-.5pt}{\drawsquare{6.5}{0.4}}}

\def\be{\begin{equation}}
\def\ee{\end{equation}}
\def\bc{\begin{center}}
\def\ec{\end{center}}
\def\bea{\begin{eqnarray}}
\def\eea{\end{eqnarray}}

\def\nn{\nonumber}

\def\det{{\rm det}}

\def\ov{\overline}
\def\qbar{\ov{q}}
\def\QQbar{\ov{Q}}

\def\tr{{\rm Tr}}

\begin{document}

\title{ Duality in $N=1$  Supersymmetric gauge theories
and recent developments}

\author{Elena Perazzi}

\address{Lawrence Berkeley National Lab, 1 Cyclotron road, Berkeley CA 94720, USA\\ 
E-mail: EPerazzi@lbl.gov}


\maketitle

\abstracts{We discuss a number of exact results in $N=1$
supersymmetric field theories. We review the results obtained by
Seiberg in Super-Yang-Mills (SYM) theories with matter in fundamental
representation.  We then consider Kutasov-type SYM theories, which
also contain matter in the adjoint representation and an appropriate
tree--level superpotential.  We finally focus on one particular case
in the latter theories, a generalization of the theories with equal
number of flavors and colors studied by Seiberg, in which non--trivial
superconformal theories appear at certain sections of the
quantum--modified moduli space.  Throughout the paper we stress the 
role played  by duality in the search for exact results.}

\section{The context}

Gauge theories are a key ingredient in our understanding of
fundamental  interactions.
While with perturbative methods we can  have a satisfactory control on 
weakly coupled gauge theories 
such as e.g. 
the theory of electroweak interactions,
strongly coupled theories such as QCD require some insight into 
non--perturbative phenomena. 

The most basic information about gauge theory dynamics
is the way gauge symmetry is realized in the vacuum state.
First of all, the gauge symmetry can be broken or unbroken in the vacuum state.
If it is broken, the gauge bosons get mass and the potential at large
distance is zero up to a constant. This is the {\bf Higgs phase}.
If it is unbroken, there are roughly speaking two cases:
\begin{itemize}
\item The vector bosons are massless and mediate long range interactions.
At large distance such interactions  give  a potential
$V(R)\sim {\alpha \over R}$. The coupling constant $\alpha$ runs
at quantum level; if  it decreases (increases) logarithmically, 
$\alpha \sim {1 \over {\rm log}(R\Lambda)}$ 
($\alpha \sim {\rm log}(R\Lambda)$, 
the theory is in the {\bf free electric 
(free magnetic) phase}. If it reaches an IR fixed point
the theory is in the {\bf Coulomb phase}.

\item Color sources are bound into singlets. This is the {\bf confining phase}.
\end {itemize}

According to the original idea of 't Hooft, there are {\em duality relations}
among the above phases. Such relations are well understood in the
case of Abelian gauge theories.
In the latter, there are both electric and magnetic charges
(think e.g. of the Georgi-Glashow model), with dual coupling strength.
The Higgs phase, where the electric charges condense
and there are solitonic magnetic flux tubes, is understood
to be dual to the confining phase, where the magnetic charges  
condense and the electric ones  are confined.
The final goal would be to understand confinement in strongly-coupled 
non-Abelian theories such as QCD, but this still requires a lot of work.

Non-Abelian theories do not contain  magnetic charges, 
at least as physical states.
Another difference with Abelian theories is that 
whenever there is matter in fundamental representation
there is no invariant distinction between Higgs and
confining phase. 
However, great advances have been made in the mid--Nineties
in {\em supersymmetric} non-Abelian gauge theories, due essentially
to the work of Seiberg and collaborators (see Ref.~\cite{dual}
and references therein).

Supersymmetric theories are more tractable than ordinary ones, due
essentially to the so--called ``power of holomorphy''.
Indeed, it is often the case that holomorphy,
together with  global symmetry considerations
and some known limits of the superpotential,  allow to determine the 
latter  completely.

The first step is to find the light degrees of freedom, i.e. the {\em 
moduli space}.
It is easy to find the classical moduli space  as the space of configurations
solving the (classical) $F$--flatness and $D$--flatness conditions.
According to a general theorem, the solutions to the
$D$--flatness conditions can be parametrized by the set of 
{\em chiral gauge invariant composites} of the theory in consideration.
Namely, in the absence of superpotential, the number of 
such composites coincides with the number of light degrees of freedom.

In the presence of a superpotential we have further conditions
(the $F$--flatness conditions) that reduce in general the 
dimensionality of the moduli space.

In the following we will consider the first examples of theories studied
by Seiberg and collaborators \cite{dual}, namely Super Yang--Mills
(SYM) theories with matter in
fundamental (and antifundamental) representation.

After that, we will consider another class of theories, studied originally 
by Kutasov \cite{kut1} and further discussed by Kutasov himself 
with Schwimmer and Seiberg \cite{kut2,kut3}.
In the context of the latter class of theories we will discuss 
a particular case, studied by the author with H.Murayama \cite{mp}.  

\section{Exact results in SYM theories and Seiberg dualities}

Let us consider first SYM $SU(N)$ theories with $F$ fundamentals
(and the same number of  antifundamentals, so that the gauge symmetry is anomaly--free).
At the classical level, these theories have a set a global symmetries, namely
\be
SU(F)_L\times SU(N)_R\times U(1)_A\times U(1)_B \times U(1)_R
\ee
under which the matter fields  transform as 
\be
Q\sim(\Yfund,1,1,1,{F-N \over F}),\ \ \ \ov{Q}\sim(1,\ov{\Yfund},1,-1,{F-N \over F}).
\ee
$U(1)_R$ an $R$-symmetry, and the charges stated above 
are those of the scalar components of $Q$, $\QQbar$.
The fermion components $\psi$ and the gauginos $\lambda$
 have charges $R(\psi)=R(Q)-1$ and $R(\lambda)=+1$, respectively.

For $F<N$ the only chiral gauge-invariant
composites are the ($F\times F$) mesons $M_{i\ov{j}}=Q_i\QQbar_{\ov{j}}$.
For $F\ge N$ there are also baryons $B^{i_1...i_{F-N}}=
\epsilon^{i_1...i_F}Q_{i_{F-N+1}}....Q_{i_{F}}$ 
and antibaryons  $\ov{B}^{\ov{i}_1...\ov{i}_{F-N}}=
\epsilon^{\ov{i}_1...\ov{i}_F}\ov{Q}_{\ov{i}_{F-N+1}}....\ov{Q}_{\ov{i}_F}$.
The gauge invariant description of the moduli space is given by the
expectation values of these operators, which, for
$F\ge N$, are subject to some classical constraints.
The rank of $M$ is at most $N$, and that of $B$, $\ov{B}$ is
at most one. If the rank of $M$ is $N$, the rank of $B$ and $\ov{B}$ is one,
and the product of the eigenvalues of $M$ is equal to the product of 
the non--zero baryon and antibaryon.
As a particular case, for $N=F$ one has the classical condition 
det$M=B \ov {B}$. 

We summarize below the phase structure of these theories, depending
on the relations between $N$ and $F$.
The easiest case is when {\bf $F>3N$}. In this case the $\beta$-function
of the theory,
\be
\beta(g)=-{b_0\over (4\pi)^2} g^3,\ \ 
b_0=3N-F
\ee
is positive, and the theory is IR free, 
i.e. it is in the free electric phase.
 
For {\bf $F<N$} there is a quantum generated superpotential, 
\be
W_{quant}=(N-F) \left({\Lambda^{3N-F}\over \det \QQbar Q}\right)^{{1\over (N-F)}}.
\ee
This can be seen as follows: this superpotential (up to a multiplicative constant)
is the only one which is compatible with all the global symmetries.
Moreover, for $F=N-1$ it is generated by instantons, as has been computed explicitly.
By giving a mass and decoupling some flavors one sees that this is indeed
the correct superpotential also for a lower number of flavors.
 
 For   {\bf ${3 \over 2} N < F < 3N$}, the theory has been argued
to possess a magnetic dual. 
The dual magnetic theory has the same number $F$ of flavors,
number of colors $\tilde{N}=F-N$, $F\times F$ singlets $M_{i \ov{j}}$ 
and a tree-level superpotential
\be
\label{treepot}
W={1 \over \mu}M_{i \ov{j}} q^i \ov{q}^j
\ee
where $\mu$ is an additional matching scale between the electric and the magnetic theory. Notice that the global symmetry group of the magnetic theory matches that of the electric theory.

Let's spend a few words on this non-Abelian duality.
In the $SU(N)$ (electric) theory with $F$ flavors, the baryons have $F-N$ indices.
Analogous degrees of freedom can be constructed as composites of $\tilde N=F-N$ states,
which we will call $q$.
The idea is to view these new degrees of freedom $q$ as physical asymptotic states,
bound together by a gauge group $SU(\tilde{N})$.
In order for the magnetic baryons to have the same global quantum numbers
as the electric ones, one needs to give the magnetic quarks the following 
transformation properties under $SU(F)\times SU(F)\times U(1)_A\times U(1)_B
\times U(1)_R$:
\be
q\sim( \ov{\Yfund}, 1, {N \over F-N}, {N\over F});\ \ \ov{q}\sim( 1, \Yfund,- {N \over F-N}, {N\over F});
\ee

However, with these quantum numbers, mesons constructed out of  $q,\ \qbar$
do not match  the electric ones.
The only way is to add in the magnetic theory $F\times F$ gauge singlets 
transforming like the mesons of the electric theory and get rid of the
$\qbar q$ composites through a tree level superpotential such as that 
of eq.(\ref{treepot}).

For this range of $F$, both the electric and magnetic theory flow
at large distance to a non--trivial fixed point, i.e. the infrared theory is
an interacting superconformal theory.

The duality relation between the electric and the magnetic theories,
originally conjectured for the range ${3 \over 2} N < F < 3N$,
can be extended beyond this range. By adding appropriate
mass deformations, the theory flows to the case $F< {3 \over 2} N$.
In the range (if any) $ N+1 < F < {3 \over 2} N$, the $\beta$--function of
the magnetic theory is positive; this is the free magnetic phase.
The flow from the range $F>3N$ to the range 
$ N+1 < F < {3 \over 2} N$ clearly shows how duality exchanges strong coupling
and weak coupling.

A new interesting behavior appears for $F=N+1$.
Here, the low energy theory shows confinement {\em without chiral symmetry breaking}. The quantum moduli space coincides with the classical one, 
including in particular the origin, where chiral symmetry is preserved.
Only the interpretation of the classical and quantum moduli space 
is different: the quantum one is understood in terms of meson
and baryon degrees of freedom, rather than in terms of quarks.
Mesons and baryons are subjected  to the  confining superpotential
\be
\label{confining}
W_{conf}={1 \over \Lambda^{2N-1}}\ (M_{i\ov{j}}B^i \ov{B}^{\ov{j}}-\det M),
\ee
where $\Lambda$ is the dimensional transmutation scale.

It is interesting to note that the above confining superpotential
can be explicitly obtained in the magnetic theory as an instanton
effect. One has to consider  the case $F=N+2$ and then decouple one flavor.
In the corresponding magnetic theory, the original gauge group
$SU(2)$ is completely broken and one has to add instanton effects 
as matching conditions in the low--energy theory.
The instanton superpotential
reproduces exactly that of eq.(\ref{confining}).

Finally, in the case $F=N$ the theory shows {\em confinement with chiral symmetry breaking}: the quantum moduli space does not coincide with the classical one, and in particular the origin, where chiral symmetry is preserved, is removed from the quantum moduli space.
The latter is characterized by the following {\em  quantum modified
constraint}:
\be
\label{1qmc}
\det M-B\ov{B}=\Lambda^{3N-F}
\ee
Note that the above constraint 
is obtained when one adds a mass term and decouples one
flavor to the superpotential of eq.(\ref{confining}).

\section{SYM theories with adjoint matter and Kutasov dualities}

In this section, we review Kutasov--type theories.
Consider a $SU(N) $ gauge theory with 
$F$ flavors in fundamental ($Q$) and anti-fundamental ($\QQbar$) 
representation and an adjoint ($X$), and with a superpotential
\be
\label{superpot}
W=\frac{h}{k+1} \ \tr X^{k+1},
\ee
where $h$ is a coupling constant of dimension $k-2$. 
The global symmetry is
$SU(F)\times SU(F)\times U(1)_B\times U(1)_R$, and the matter transformation
properties are:
\bea
&&Q\sim (\Yfund, 1,1,1-{2\over k+1}{N\over F}),\ \ 
\QQbar\sim (1, \ov{\Yfund},-1,1-{2\over k+1}{N\over F}),\nn\\
&&X \sim(1,1,0,{2 \over k+1}).
\eea 
The $F$-flatness condition from (\ref{superpot}) are:
\be
\label{f-flat}
X^k-{1 \over N} {\rm Tr} X^k=0.
\ee
Up to complexified gauge transformations\footnote{Remember the general theorem
according to which solving the $F$--flatness conditions and modding out by complexified gauge transformations is equivalent to solving both the $F$--flatness
and the $D$--flatness and modding out by ordinary gauge transformations.}, 
there are two kinds
of solutions to the above equation:
either $X$ is diagonal and
it is $X^k=v^k I$, where $v$ is an arbitrary complex number and $I$ is
the identity matrix, or it has all zero diagonal entries (and thus it
is a singular matrix with vanishing $k$th power) 
We can think
of the latter case as the origin of the flat direction $v$ in moduli
space.

For $v\neq 0$, vacua of the gauge theory can be labeled by 
sequences of integers $(r_1,r_2,...,r_k)$, with $\sum r_i=N$,
where $r_i$ is the number of eigenvalues of $X$ residing 
in the $i-$th of the $k$ roots of $v^k$.
The gauge group is broken by the $X$ expectation value to
\be
\label{elfact}
SU(N) \rightarrow SU(r_1)\times SU(r_2)\times ... \times SU(r_k)\times
U(1)^{k-1}.  
\ee   
Thus, at low energies we are left with $k$ decoupled SQCD theories with
gauged baryon number(s), whose behavior has been described in the previous section.
  
As an extension of the Seiberg duality conjecture discussed above, 
Refs.~\cite{kut1,kut2,kut3} suggested that the
theory under discussion is dual to a theory with
$SU(kF-r)$ gauge group, $F$ flavors of (dual) quarks ($q$) and
antiquarks ($\qbar$), $k$ singlets $M_j$ and an adjoint $Y$.
The magnetic theory has the same global symmetry group as the electric one,
$SU(F)\times SU(F)\times U(1)_B\times U(1)_R$,  
and the matter transformation properties are:
\bea
&&q \sim (\ov{\Yfund},1, {N\over F-N},1-{2(kF-N)\over (k+1) F}),
\ \qbar \sim (1,\Yfund, -{N\over F-N},1-{2(kF-N)\over (k+1) F}),\nn\\
&&Y\sim(1,1,0,{2\over (k+1)})\nn\\
&& M_j\sim(\Yfund,\ov{\Yfund},0,2-{4N\over (k+1)F}+{2\over (k+1)}(j-1)).
\eea
Except for the case $k=2$, $2F-N=2$, 
the magnetic tree-level superpotential is taken to be
\be
\label{magn}
W_{magn}=-\frac{h}{k+1} {\rm Tr}Y^{k+1} 
+ {h \over \mu^2} \sum_i M_i \qbar Y^{k-i} q.
\ee
The $F$-flatness condition for $Y$ is similar in form to that of
eq.(\ref{f-flat}), therefore the magnetic moduli space also contains
a flat direction $v$, analogous to that of the electric theory.
Points in the electrical moduli space where the $SU(N)$ theory splits
into the product of the $k$ $SU(r_i)$ theories correspond to points in
the magnetic moduli space where the magnetic $SU(kF-N)$ theory splits
into the product of the corresponding dual $SU(F-r_i)$ theories.
Notice that points in the classical electric moduli space
with some $r_i<F$ are removed from the corresponding
magnetic moduli space because $SU(F-r_i)$ cannot exist then. Thus, the
two spaces do not agree classically, but only quantum mechanically.

In the case $k=2$, $2F-N=2$, Tr$Y^3=0$, and, as shown in Ref.\cite{mp},
the magnetic superpotential requires an additional term, and is thus
\be
\label{additional}  
W_{magn}={h \over \mu^2} \sum_i M_i \qbar Y^{k-i} q
-\left({{\rm det} M^{(1)}\over  \Lambda^{(1)\ 3N/2-(F+1)}}+
{{\rm det} M^{(2)}\over  \Lambda^{(2)\ 3N/2-(F+1)}}\right),
\ee
where 
\be
M^{(1),(2)}= {M_1\pm v  M_2 \over 2v},
\ee
and, consistently with the general case, $v=\sqrt{{1 \over 2 }{\rm Tr} Y^2}$.
  
At the origin $v=0$, the adjoint doesn't
decouple from the low-energy theory
and the moduli space can be described at all energies by generalized mesons
\be
\label{genmes}
(M_j)^i_{\ov{i}}=\QQbar_{\ov{i}} X^{j-1}Q^i;\ \ \ j=1,...,k;\ \
i,\ov{i}=1,...,F\ , 
\ee
baryons
\be
\label{genbar}
B^{(n_1,n_2,...,n_k)}=Q^{n_1}(XQ)^{n_2}....(X^{k-1}Q)^{n_k}
\ee
and, finally, Tr$X^j$ with $j=1,...,k$.
The mesons (\ref{genmes}) can be also thought of as blocks of the matrix
\be
\label{mes1}
\left(\begin{array}{ccccc}
       \QQbar Q & \QQbar XQ  &...  & \QQbar X^{k-2} Q  & \QQbar X^{k-1} Q \\
       \QQbar X Q  & \QQbar X^2 Q  &...  & \QQbar X^{k-1} Q      & 0  \\
        .   &     &   &    &        \\
        .   &     &   &    &        \\
        .   &     &   &    &         \\
       \QQbar X^{k-1}Q   & 0   & ... & 0 & 0    
\end{array} \right)
\ee
constructed from the ``dressed'' quarks and anti-quarks
\be
\label{dressed}
Q_{(l)}=X^{(l-1)}Q;\ \ \QQbar_{(l)}=X^{(l-1)}\QQbar;\ \ l=1,...k\ \ .
\ee 

A mapping between the above gauge-invariant operators and those of the
magnetic dual can be established as follows: the mesons in
eq.(\ref{genmes}) correspond to the elementary singlets of the
magnetic theory. The correspondence between the electric baryons ($B$)
defined in eq.(\ref{genbar}) and magnetic ones ($b$), constructed in
an analogous way out of $q$ and $Y$, is
\be
B^{(n_1,n_2,....,n_k)} \leftrightarrow  b^{(m_1,m_2,....,m_k)},\ \ \
m_l=F-n_{k+1-l},\ \ l=1,...,k 
\ee
and the traces Tr$X^j$ are simply mapped to the analogous $-$Tr$Y^j$.

As for ordinary Seiberg duality, non--trivial tests of this
new conjectured duality include 't Hooft anomaly matching, 
matching of the electric and magnetic moduli space
and the fact that duality is preserved under mass deformations.

The discussion in Refs.~\cite{kut1,kut2,kut3} established a picture of
the IR behavior of the theory at $v=0$ for all the values of $F$ such
that $kF-N>1$: the theory is in the free electric phase for $F>2N$, in
the free magnetic case for $F<{2\over 2k-1} N$ and in the non-Abelian
Coulomb phase for the values of $F$ in the intermediate range.

Furthermore,  Csaki and Murayama studied the case
$kF-N=1$ \cite{cm}, by adding a mass deformation to the case
$kF-N=k+1$. They found that for $kF-N=1$ the theory is always
confining, and obtained explicitly the confining superpotential as a
$k$-instanton effect in the magnetic theory.

In Ref.\cite{mp} it was found that 
duality considerations can also
elucidate the behavior of the theory for $kF=N$.

\section{Kutasov theories with {\bf $kF=N$}}

Consider the $SU(kF)$ theory with $F$ pairs of quarks $Q$, 
$\ov{Q}$ and an adjoint field $X$ with the superpotential of
eq.(\ref{superpot}). 

The classical moduli space is given in terms of the mesons $M_{j} = \ov{Q} 
X^{j-1} Q$ $(j = 1, 2, \cdots, k)$, baryons $B = Q^{F} 
(XQ)^{F} \cdots (X^{k-1} Q)^{F}$, $\ov{B} = \ov{Q}^{F} 
(X\ov{Q})^{F} \cdots (X^{k-1} \ov{Q})^{F}$, ${\rm Tr}X^{j}$, $j=2,3,...,k$.
If we define the matrix ${\cal M}$ such that ${\cal M}_{ij}=M_{i+j-1}$
for $i+j\le k+1$ and ${\cal M}_{ij}=v^k M_{i+j-(k+1)}$ otherwise,
baryons and mesons are subject to the constraint
\begin{equation}
\label{constraint}
    \det {\cal M} - B\ov{B}=0.
\end{equation}
In the limit $v\rightarrow 0$ ${\cal M}$ reduces to the matrix in 
eq.(\ref{mes1}) and the constraint (\ref{constraint}) 
simplifies to
\be
\label{constr1}
(-1)^{{k(k-1)\over 2}} {\rm det} (M_k)^k -B\ov{B}=0.
\ee

Along the $F$-flat and $D$-flat direction $X^{k} = v^{k} I$ with $v
\neq 0$, $X$ takes the form $X = v\, {\rm diag}(\underbrace{1, \cdots,
  1}_{n_{1}}, \underbrace{\omega, \cdots, \omega}_{n_{2}}, \cdots,
\underbrace{\omega^{k-1}, \cdots \omega^{k-1}}_{n_{k}})$ with $\omega
= e^{2\pi i/k}$.  It is easy to see, however, that only the choice
$n_{1} = \cdots = n_{k} = F$ is left on the quantum moduli space.  For
any other choice at least one of the remaining $SU(n_{j})$ gauge
groups satisfies $n_{j} > F$, and hence a dynamical superpotential is
generated and the moduli space is lifted quantum mechanically.  For
the only possible choice $n_{1} = \cdots n_{k} = F$, the gauge group
is broken to $SU(F)^{k} \times U(1)^{k-1}$. Each $SU(F)$ factor
has $F$ flavors, thus each subsector is characterized by a quantum modified 
constraint of the form of eq.(\ref{1qmc}). 
More precisely, after finding the correct dynamical scale for
each of the subsectors, one finds the following quantum modified constraints:
\be
    ({\rm det} M^{(j)} - B^{(j)}\ov{B}^{(j)}-\Lambda^{(j)2F})
    =  
    \left({\rm det} M^{(j)} - B^{(j)}\ov{B}^{(j)}
    - \frac{h^{F}\Lambda^{2kF-F}}{k^F(v\omega^{j-1})^{(k-1)F}}\right),
    \label{eq:kmoduli}
\end{equation}

Now we must take the limit  limit $v \rightarrow 0$.  Clearly,
the moduli space described by Eq.~(\ref{eq:kmoduli}) is singular as $v
\rightarrow 0$.  We approach this limit in two different ways.

The first method is to approach $v \rightarrow 0$ when the
$U(1)^{k-1}$ factors are always broken, i.e. when the product of baryon and antibaryon
is non--zero for all the subsectors.  
In this case, by  appropriately relating the degrees of freedom
of the subsectors to those of the high-energy theory,
one finds the limit:
\begin{equation}
    B\ov{B} =  \prod_{j} (-1)^{{(k-1)F\over 2}}\left( {\rm det} M_{k}
    - h^{F}\Lambda^{2kF-F} \right).
\end{equation}
We again stress that this quantum modified constraint is found 
{\it as long as}\/ you approach the origin with all
the gauge groups always completely broken.

The magnetic dual of the $SU(kF)$ theory considered in this section
is  defined as the low energy theory obtained
after adding an appropriate mass deformation to the $SU(k)$ magnetic theory with $F+1$ flavors.
In such dual theory the gauge group is completely broken and,  
as shown in detail in Ref.\cite{mp}, the above quantum--modified constraint is obtained as an instanton effect, {\em when $k\neq 2$}.
For $k=2$, the superpotential piece proportional to 
Tr$Y^3$ vanishes in the magnetic $SU(2)$ theory, and 
instanton effects are also vanishing. 
The above constraint is however reproduced thanks to the additional superpotential term of Eq.(\ref{additional}).

What about taking the limit $v \rightarrow 0$ while  keeping some or 
all of $U(1)$'s unbroken?  There are many reasons to believe that 
this limit leads to an interacting superconformal theory.  One way to 
see it is as follows.  We can always force the baryons to vanish in 
Eq.~(\ref{eq:kmoduli}), by adding a mass term to the quarks.  By 
adding a common mass term for simplicity,
\begin{equation}
    W = \sum_{j=1}^{k} X_{j} 
    \left({\rm det} M^{(j)} - B^{(j)}\ov{B}^{(j)}
    - \frac{h^{F}\Lambda^{2kF-F}}{k^F (v\omega^{j-1})^{(k-1)F}}\right)
    + m {\rm Tr} M_{1},
\end{equation}
and noting
\begin{equation}
    M_{1} = \sum_{j=1}^{k}  M^{(j)},
\end{equation}
we can solve $\partial W/\partial M^{(j)} = 0$ to find that $X_{j} 
\neq 0$.  This is enough to force all baryons to vanish.  Then we can 
ask the question what happens in the $v\rightarrow 0$ limit.  Because 
quarks are massive, we can integrate them out first instead, and add 
the superpotential $h {\rm Tr}X^{k+1}$ afterwards.  Once the quarks 
are integrated out, the theory is nothing but the $N=2$ Yang--Mills 
theory, whose curve is known.  Adding the superpotential is known to 
make the theory flow to an Argyres--Douglas fixed point.  It was 
worked out explicitly for the $SU(3)$ and $k=2$ case 
\cite{ad}, but it is believed that any $SU(N)$ theory with 
any $k$ would lead to such non-trivial fixed-point theories, as long 
as $k < N$.  Therefore for $F \geq 2$, the theory will flow to 
superconformal theories.  When $F=1$, however, the superpotential is 
(presumably) irrelevant, and the theory is given by the Coulomb branch 
of the entire $SU(N)$ $N=2$ Yang--Mills.  

In the magnetic theory, the quantum modified constraint is
also satisfied when $q=\qbar=Y=0$ and only the mesons get a vev, in
which case the SU(k) gauge symmetry is unbroken.  At first sight, it
is not obvious that this point belongs to the moduli space.  The
instanton superpotential is generated (for $k\neq 2$) 
if the gauge group is broken; if
it was not, the $F$-flatness condition would be the classical one,
which is not satisfied by $q=\qbar=Y=0$.  On the other hand this point
can be reached from the direction $b^{(i)}=\ov{b^{(i)}}=0$ in the
limit $v\rightarrow 0$ and from the direction $v=0$, ${\rm det}
M_k=h^F\Lambda^{2N-F}$, $B(\ov{B})=0$, $\ov{B}(B)\rightarrow 0$ on the
moduli space.  In this limit, $U(1)^{k-1}$ gauge invariance is
unbroken and additional charged massless fields can arise at
singularities on the moduli space where $SU(k)$ is recovered
classically.  Therefore we conclude that this limit is on the moduli
space, where the theory becomes superconformal.  In the case $k=2$,
this is also in agreement with the results obtained in
Ref.~\cite{isphases} for the particular case of an $SU(2)$ theory with
two doublets and a triplet.

\end{document}